 \documentclass[aps,prb,preprint,superscriptaddress,showpacs]{revtex4-1}

 \usepackage{graphicx,psfrag}
 \usepackage[latin1]{inputenc}
 \usepackage{natbib}
 \usepackage{amsmath, amsthm, amssymb}
 \usepackage{color}
 \usepackage{comment}

 \newcommand{\notes}[1]{}

 \newcommand{\beq}{\begin{equation}}
 \newcommand{\eeq}{\end{equation}}
 \newcommand{\beqnn}{\begin{equation*}}
 \newcommand{\eeqnn}{\end{equation*}}
 \newcommand{\beqas}{\begin{eqnarray*}}
 \newcommand{\eeqas}{\end{eqnarray*}}
 \newcommand{\beqa}{\begin{eqnarray}}
 \newcommand{\eeqa}{\end{eqnarray}}

\begin{document}

\title{Effect of temperature on spin-transfer torque induced magnetic solitons}
\date{\today}
\author{Sergi Lend\'inez}
\altaffiliation[Currently at: ]{Materials Science Division, Argonne National Laboratory, Lemont IL 60439, USA.}
\affiliation{Department of Condensed Matter Physics, University of Barcelona, 08028 Barcelona, Spain}
\author{Jinting Hang}
\affiliation{Department of Physics, New York University, New York, New York 10003, USA}
\author{Sa\"ul Vélez}
\affiliation{CIC nanoGUNE, 20018 Donostia-San Sebastian, Basque Country, Spain}
\author{Joan Manel Hern\`andez}
\affiliation{Department of Condensed Matter Physics, University of Barcelona, 08028 Barcelona, Spain}
\author{Dirk Backes}
\altaffiliation[Currently at: ]{Department of Physics, Loughborough University, Loughborough LE11 3TU, UK}
\affiliation{Department of Physics, New York University, New York, New York 10003, USA}0
\author{Andrew D. Kent}
\affiliation{Department of Physics, New York University, New York, New York 10003, USA}
\author{Ferran Maci\`a}
\affiliation{Department of Condensed Matter Physics, University of Barcelona, 08028 Barcelona, Spain}
\affiliation{Institut de Ci\`encia de Materials de Barcelona (ICMAB-CSIC), Campus UAB, 08193 Bellaterra, Spain}

\begin{abstract}

Spin-transfer torques in a nanocontact to an extended magnetic film can create spin waves that condense to form dissipative droplet solitons. Here we report an experimental study of the temperature dependence of the current and applied field thresholds for droplet soliton formation, as well as the nanocontact's electrical characteristics associated with droplet dynamics. Nucleation of droplet solitons requires higher current densities at higher temperatures, in contrast to typical spin-transfer torque induced switching between static magnetic states. Magnetoresistance and electrical noise measurements show that soliton instabilities become more pronounced with increasing temperature. These results are of fundamental interest in understanding the influence of thermal noise on droplet solitons, and in controlling their dynamics.

\end{abstract}

\maketitle


The spin transfer torque (STT) effect between itinerant electron spins and magnetization \cite{Slonczewski1996,Berger1996,Katine2000} is an important discovery in nanomagnetism because it provides a new means of manipulating magnetization states without using magnetic fields. The angular momentum of a spin-polarized electrical current can be transferred to the magnetic moments of a magnetic material \cite{Ralph2008}. An important application of STT is magnetic random access memory (MRAM) that uses the static states of bistable nanomagnets whose magnetization is oriented using spin-polarized currents\cite{Kent2015}. The physics governing transitions between static magnetic states under the STT effect in bistable nanomagnets, such as those incorporated in magnetic tunnel junction (MTJ) pillars, are typically described through statistical mechanics \cite{Coffey_JAP_2012} considering thermal fluctuations and effective energy barriers that depend on spin-polarized current. An increase in temperature leads to a faster rate of transitions between magnetic states and thus to a reduced current density for STT switching. The STT effect can also serve to create or modify dynamic collective excitations such as spin waves or solitons\cite{Brataas2012}. The stability of these collective spin excitations is, in general, more complex than magnetic static states such as spin valves or localized magnetic domains because besides the competing magnetic energies, dissipation also plays a role. 

Experimental studies have used the localized spin currents in a nanometer-scale electrical point contact to an extended ferromagnetic thin film to excite linear propagating modes \cite{Demidov2010,Madami2011} and soliton modes \cite{Demidov2012,Backes2015,Bonetti2015}. \emph{Dissipative magnetic droplet solitons} (droplets hereafter) are nonlinear confined wave excitations consisting of partially reversed precessing spins that can be created in films with perpendicular magnetic anisotropy (PMA) through the local suppression of the magnetic damping \cite{Hoefer2010}. Droplets have been experimentally created using the STT effect in electric nanocontacts to PMA films \cite{Mohseni2013, Chung2014, Mohseni2014, Macia2014, Lendinez2015, Chung2015, Akerman2016NatComm}. Recently reported experiments have shown that the stability of these collective excitations is limited by the appearance of drift instabilities, which were attributed to the disorder---local variations of the effective magnetic field \cite{Lendinez2015,Bookman2015}. Wills \emph{et al.} \cite{Hoefer2016} identified theoretically an intrinsic deterministic linear instability and thermal fluctuations as additional instability mechanisms.


In this study we report temperature-dependent measurements of the thresholds for spin polarized current induced excitation of droplet solitons. We have mapped the conditions in applied field and electrical current that allow droplet-soliton formation as a function of temperature. We have found that increasing temperature destabilizes droplet solitons and require larger current densities to create and sustain them. Additionally, we have observed that temperature enhances electrical noise associated with droplet-soliton dynamics.
\\

\begin{figure}[htb!]
\includegraphics[width=0.9\columnwidth]{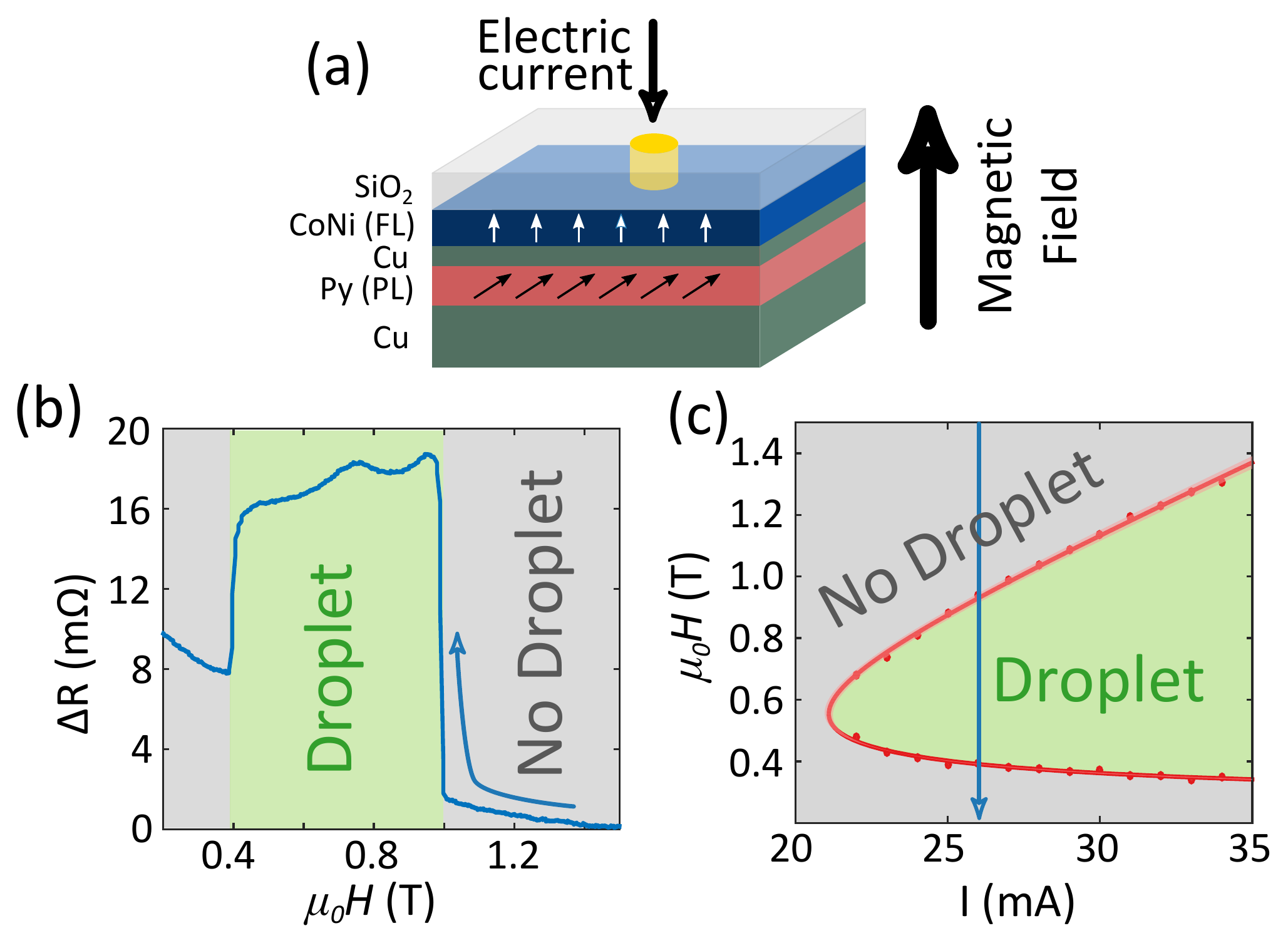}
\caption{(a) Schematic of the electrical point contact to a magnetic bilayer structure. An electrical current flows through a nanocontact to a thin ferromagnetic layer (the free layer, FL) and a spin-polarizing layer (PL). (b) Nanocontact resistance variation as a function of the field at $T=150$ K with a current of 27 mA, showing the creation and annihilation of a droplet state. The green area represents the droplet state, whereas in the gray area, the FL magnetization is aligned with the external applied field. (c) Measured stability map for the droplet soliton as a function of magnetic field and electrical current at 150 K. The blue vertical line indicates the current at which the field sweep presented in (b) is conducted.}
\label{fig1}
\end{figure}

The samples we studied were metallic nanocontacts (150 nm in nominal diameter) to a magnetic bilayer structure consisting of a free layer (FL) composed of Cobalt (Co) and Nickel (Ni) with PMA and a polarizing layer (PL) of Ni$_{80}$Fe$_{20}$ (permalloy) with in-plane magnetization \cite{Macia2014} [see, Fig.\ \ref{fig1}(a)]. The FL and PL were magnetically decoupled with a 10-nm-thick Copper (Cu) layer. We have characterized the layers using magnetometry and ferromagnetic resonance spectroscopy\cite{Macia2012}. The FL has an effective anisotropy field of $\mu_0 H_{\text{eff}}=\mu_0(H_K-M_s)\simeq 0.25$~T, indicating a strong PMA. The PL is a soft magnetic material with a saturation magnetization of $\mu_0M_s\simeq1$~T and its magnetization lies in the plane in the absence of an applied field.



The electrical response of the nanocontact depends on the relative orientation of the FL and the PL's magnetization due to the giant magnetoresistance (GMR) effect. A low resistance value corresponds to the layer magnetizations being parallel (P), whereas a high value corresponds to an antiparallel (AP) alignment; the overall normalized magnetoresistance is thus $\overline{R}_0=(R_\mathrm{AP}-R_\mathrm{P})/R_P$, where $R_\mathrm{AP,P}$ is the resistance of the nanocontact with AP and P magnetization states. At large perpendicular applied fields, $H> M_{s}$, the PL magnetization is saturated in the field direction, resulting in a P state with a minimum value for the resistance of the nanocontact. When a droplet state forms in the FL, the magnetization (partially) reverses and an increase of the resistance is expected. The resistance change will be the largest, $\overline{R}_0$, when the spins in the droplet are completely reversed and the droplet fully occupies the nanocontact region. At small applied fields, $H<M_S$, the PL magnetization has only a small component in the direction of the FL magnetization, which results in a linear decrease of the nanocontact resistance with the applied field; droplet formation would still produce a change in resistance, this time, proportional to the magnetization component of the PL in the perpendicular direction. In Fig.\ \ref{fig1}(b) we plot a measured magnetoresistance curve \cite{Mohseni2013, Macia2014, Mohseni2014} at a fixed current of 27 mA ramping the perpendicular applied field from high to low field at a fixed temperature of 150 K. This shows first the creation (1 T) and then the annihilation of the droplet state (0.4 T). Figure \ref{fig1}(c) shows a measured droplet stability map in magnetic field and current space \cite{Lendinez2015, Akerman2016NatComm} build from magnetoresistance curves measures at different applied current values and a fixed temperature of 150 K. The non-soliton state is plotted in gray and corresponds to a lower resistance state whereas the the green area represents the droplet state, a higher resistance state.

In order to study the temperature dependence of the dc- and ac-resistance, we wire-bonded our nanocontact to sample holders capable of transmiting microwave signals up to 4 GHz and introduced them into a cryostat that allowed variations of temperature from 2 K to 350 K and bipolar fields up to 9 T. We were able to vary the angle between the film plane and the applied field by rotating the sample holder. We used a current source and a voltmeter to measure the dc-resistance response, and a spectrum analyzer to study the spectral composition of the electrical noise of the nanocontact. A bias T separated the dc and ac components of the signal. Resistance measurements allowed us to determine the relative orientation of the magnetization between the free (FL) and polarizing (PL) layers via the GMR effect. Experimental results showed in the following are done with a magnetic field applied perpendicular to the film plane.
\\



We first studied the creation and annihilation of the droplet through dc measurements of the magnetoresistance. The current was fixed to minimize temperature variations and the external field was swept from $-3$~T to 3~T and back [Fig.\ \ref{fig1}(b) shows data at $T=150$ K with decreasing applied field]. At zero applied field the PL and FL are orthogonal and current through the nanocontact that is polarized in the direction of the PL creates a torque that averages to zero over one precessional cycle. As the magnetic field increases, the PL magnetization tilts and increases the electrical current's spin polarization in the direction of the FL magnetization; eventually the STT effect on the FL compensates the damping and allows for the nucleation of a droplet state. Larger fields ultimately destabilize and annihilate the droplet, favoring an alignment of the two layer's magnetization in the direction of the applied field. As we sweep the field down back to zero, droplet states nucleate and annihilate at similar field values as in the sweep up, showing hysteresis in some cases \cite{Mohseni2013,Macia2014}, which indicate stability of droplet states (see supplemetary materials for a sample with pronounced hysteresis).

\begin{figure}[htb!]
\includegraphics[width=0.8\columnwidth]{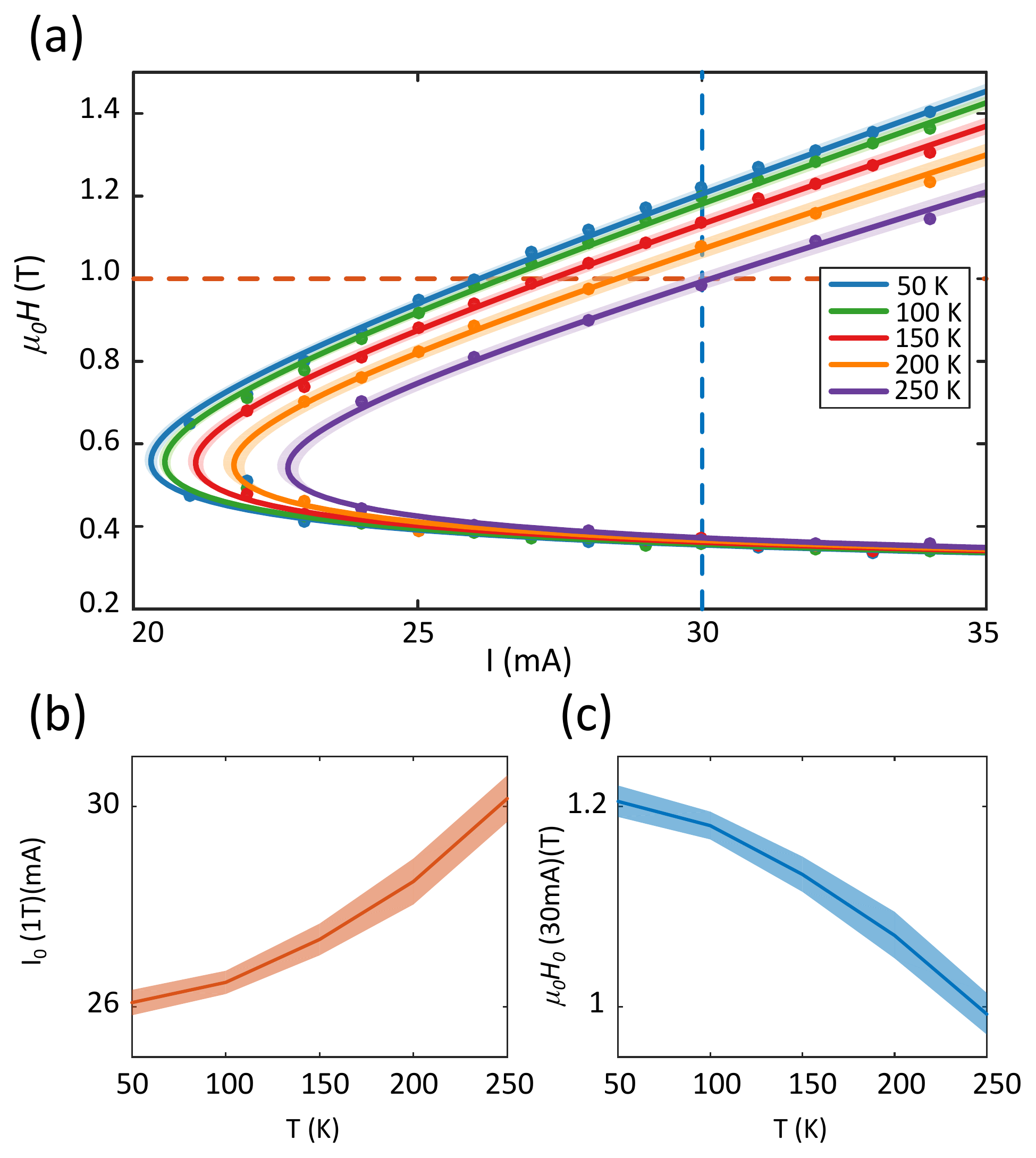}
\caption{(a) Stability maps of soliton states as a function of temperatures (50 K to 250 K). The points show the current and field combination at which the droplet is created. The curves are fits to the data. (b) and (c) Temperature dependence of the onset current at a fixed magnetic field of 1~T (b), and of the onset field at a fixed current of 30~mA (c), obtained from the fits. The light colors in the graphs show the uncertainties.}
\label{fig2}
\end{figure}

We measured the stability maps of droplet-soliton states as a function of temperature. Figure~\ref{fig2} shows field and current values that allow for droplet-state formation at different temperatures (from 50~K to 250~K). We observe how the region of droplet existence is reduced as the temperature increases, indicating that larger current densities are needed to create and sustain droplet states when the temperature increases (measurements on a different sample are presented in the supplementary materials in the range of 2 K to 350 K).

The boundaries determining the region of droplet existence present a minimum current at a particular value of applied field (below that current value the droplet cannot be generated regardless of the field). An analytic prediction for the variation of the onset current as a function of the applied field is not available because of the sample's orthogonal geometry (i.e., PL perpendicular to FL). Hoefer \emph{et al.} \cite{Hoefer2010} provided an expression for the minimum sustaining current in the case of having a PL always magnetized normal to the film plane that was proportional to the damping parameter and to the precessing frequency---that is also proportional to the applied field. However, in our samples we need an applied field larger than $\sim$1 T to ensure the PL magnetization is normal to the film plane. At fields that do not completely align the magnetization of the two layers, the degree of spin polarization of the electrical current in the direction of the FL plays an important role. The polarization degree depends on the magnetization component of the PL normal to the film plane, and the latter is proportional to the applied field. Thus the onset current has an inverse dependence with the applied magnetic field at low fields, as reported in several experimental studies \cite{Mohseni2014, Macia2014, Akerman2016NatComm, Ozyilmaz2003}. Chung \textit{et al.} \cite{Akerman2016NatComm} recently used an expression for the onset current as a function of the field that accounts for the effect of a PL that is not aligned with the FL. Their approximation for the onset current as a function of the applied field summarizes the previous discussion; its derivation is based on the Slonczewski critical current condition \cite{Slonczewski1999} in a material with PMA and consists of a term proportional to the applied field that dominates at large fields and another that is inversely proportional to the same field that dominates at lower field values,
\begin{equation}
I_\mathrm{c}=a\mu_0H+\frac{b}{\mu_0H}+c.
\label{stability}
\end{equation}
Our data fits poorly to such expression (see supplementary materials) but instead it fits well when we consider a shifted value for the applied field, $H\rightarrow (H-H_0)$, with $\mu_0H_0\simeq0.27$ T, which matches with the anisotropy field of the FL.  Figure~\ref{fig2}(a) shows the fits to the data (see supplementary materials for plots of fitting parameters, $a$, $b$, $c$, and $H_0$, as a function of temperature).

We have observed that larger currents are required to create droplet solitons at higher temperatures. As a result there is also a decrease in the maximum field at which droplet states occur. We plotted these two indicative quantities in Fig.~\ref{fig2}(b,c); the onset current with a fixed field of 1~T and the onset field (sweeping down from large fields) at a fixed current of 30~mA. Notice that the changes become more pronounced as the temperature increases and seem to saturate when it decreases.
\\


\begin{figure*}[htb]
\includegraphics[width=120mm]{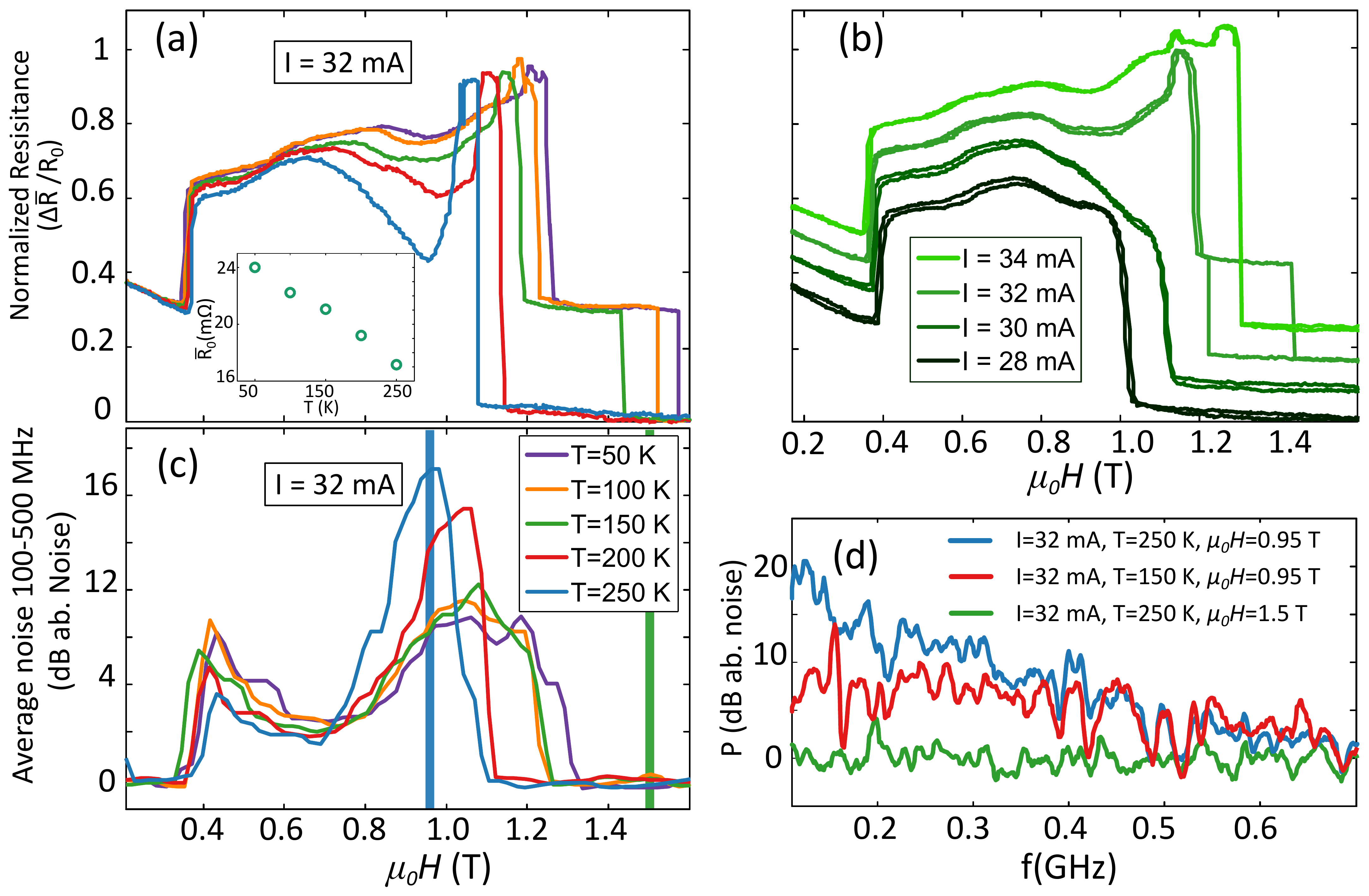}
\caption{(a) Normalized magnetoresistance as the field sweeps up at a fixed current of $I=32$ mA for temperatures ranging from 50 to 250 K. The inset shows the normalizing values, $\overline{R}_0$. (b) Magnetoresistance curves for both sweeps up and down at a fixed temperature $T=150$ K for different applied currents ranging from 28 to 34 mA. (c) Low-frequency signal above noise averaged in the range 100-500 MHz measured at the same time as the magnetoresistance curve in (a). Vertical bars indicated the magnetic fields at which spectra correspond in (c). (d) Low-frequency response measured at $I=32$ mA and $T=250$ K at 1.5 T (green curve), and at 0.9 T (blue curve), and $T=150$ K at 0.9 T (red curve).}
\label{fig3}
\end{figure*}

We characterize the overall change in resistance between P and AP states at each temperature using magnetotransport measurements at small applied currents. The sample resistance varies with temperature and the total MR changes as well. Figure\ \ref{fig3}(a) shows the normalized MR curves measured at different temperatures when the magnetic field is swept up from zero to 1.6 T. Overall resistance changes at each temperature, $\overline{R}_0(T)$, are plotted in the inset of Fig.~\ref{fig3}(a). The green curve in Fig.~\ref{fig3}(a), for instance, shows an MR curve measured at 32~mA and 150~K; we can see a step increase in resistance at $0.3$~T corresponding to the creation of the droplet state and a second step (decrease) at $1.18$~T corresponding to the droplet-state annihilation. At temperatures above 150 K the MR curves show an intermediate state between the first step decrease at $1.2$~T and a second step decrease at $1.5$~T that indicates the presence of a partially reversed magnetization state in the nanocontact.

The MR curve measured at 150 K is plotted in Fig.\ \ref{fig3}(b) along with curves measured at other applied currents but at the same temperature and we see that the onset and annihilation fields vary as in Fig.\ \ref{fig2}. We observe that, within the high-resistance states (droplet states), the resistance fluctuates with the applied field indicating different soliton configurations \cite{Macia2014} that are reproducible at different applied currents [see, Fig.\ \ref{fig3}(b) for fields between 0.4 and 0.8 T]. However, when varying the temperature we also observe, in addition to the resistance fluctuations, a decrease in the overall resistance change  with increasing the temperature. We thus confirm that the increase in temperature reduces the effective amount of reversed magnetization, either because the droplet state becomes smaller than the nanocontact or because on average it spends less time in the contact region, such as due to drift instabilities \cite{Lendinez2015, Hoefer2016}. In the supplementary materials the evaluation and the analysis of the overall change in MR in different devices with the same layer stack is shown.


In order to investigate the effect of instabilities in the droplet states as a function of temperature we have measured the electrical noise at frequencies below 4~GHz with a spectrum analyzer at the same time we measured the dc resistance. We associate the electrical noise signal with motion of the droplet soliton beneath the nanocontact; the signal is present when a droplet state forms \cite{Lendinez2015,Akerman2016NatComm}. The electrical noise signal is mostly a $1/f$-type [see, Fig.\ \ref{fig3}(d)], but in some cases is accompanied by a resonant peak at frequencies of the order of hundreds of MHz \cite{Lendinez2015}. 

We observed the appearance of a low-frequency signal (electrical noise) when the sample was in a droplet state. Figures~\ref{fig3}(a,c) show simultaneous measurements of the dc resistance and the low-frequency spectral signal, which is averaged between 100 MHz and 500 MHz at each magnetic field [see, Fig.\ \ref{fig3}(d) for frequency sweeps at a given field]. We can compare the MR curves at a fixed current of 32 mA and the corresponding low-frequency signal. At 250~K we observe that when the droplet develops at $0.35$~T there is an increase in the dc resistance and a peak in the low-frequency signal. As the field is swept up the droplet state stabilizes and we observe a decrease in the low-frequency signal (field values between $0.5$~T and $0.8$~T). As we continue with the field sweep the dc resistance slightly decreases and the spectral signal increases indicating additional dynamics in the droplet state. A further increase of the applied field eventually annihilates the droplet state. We note that the small changes in the low-frequency signal within the droplet state as a function of the applied field shown in Figs.\ \ref{fig3}(c) were reproducible every time we swept the field.

We have thus observed an evolution of droplet states, both in dc resistance and in the electrical noise, as we swept the field. There is a correlation between the power level of the low-frequency signal and the dc-resistance signal at different temperatures; droplet states giving a large dc resistance generally produce a smaller low-frequency spectral signal indicating a larger stability. However, we note that the onset of droplet solitons at low fields is always associated with high levels of electrical noise and that droplet states that only present fractional variations of the dc resistance, for instance the MR curves in Fig.\ \ref{fig3}(a) for 50~K, 100~K or 150~K above 1.2~T, do not present measurable electrical noise.
\\


In summary, we have found that higher temperatures destabilize STT-induced droplet solitons; nucleation and stabilization of droplet solitons require higher current densities at higher temperatures. We have measured the droplet stability maps in current and field at different temperatures and provided an expression for the onset current based in Chung \textit{et al.}'s \cite{Akerman2016NatComm} approximation with an additional effective field of the order of the anisotropy of the FL. We have quantified the overall change in resistance associated with droplet states in the nanocontact and found that it reduced with increasing temperature indicating a partial reversal of the FL magnetization or a smaller size of the soliton. Further, we have recorded the low-frequency spectral signal associated with drift instabilities in the droplet solitons and observed that there is a correlation between states with an incomplete reversal of the magnetization and states showing a stronger drift-resonance signal. We conclude that the STT effect is more complex in dynamic solitons, and probably other collective dynamic spin excitations, than in nanopillars \cite{Urazhdin2003}. Our results clearly show that temperature effects the formation of STT-induced droplet solitons suggesting that the interaction between STT and thermal magnons might compete with magnon condensation.

\section*{Acknowledgements}

We acknowledge technical support from Luis Hueso and Fèlix Casanova at CIC nanoGUNE. F.M. acknowledges financial support from the Ram\'on y Cajal program through RYC-2014-16515 and from MINECO through the Severo Ochoa Program for Centers of Excellence in R\&D (SEV-2015-0496). SV acknowledges support from the EU 7th FP under the ERC 257654-SPINTROS and by the Spanish MINECO through MAT2015-65159-R. Research at UB was partially supported by MINECO through MAT2015-69144. Research at NYU was supported by NSF-DMR-1610416.

\newpage
\huge{Supplemental material}
\normalsize

\section*{Data on different samples}
Here we present data on a different sample fabricated on the same layer stack with nominally identical geometry. The sample has been measured in a larger temperature range although we did not measure the low-frequency signal (and this is the reason that this data is not presented in the main manuscript).
\begin{figure}[htb!]
	\includegraphics[width=0.8\columnwidth]{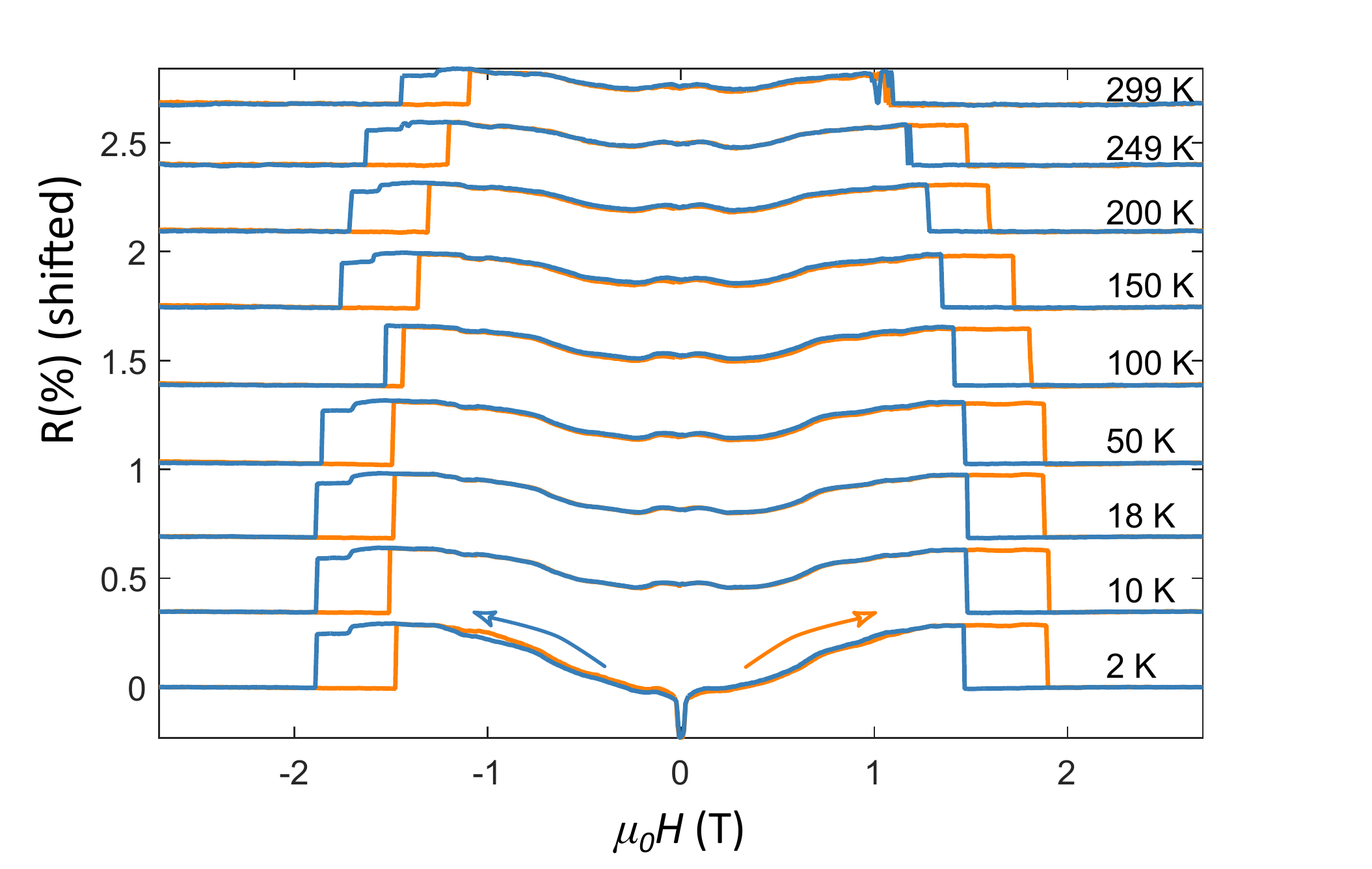}
	\caption{\small{MR measurements as the field is swept up from -3 T to 3 T
			(orange) and down from 3 T to -3 T (blue), for different temperatures in the range 2 K - 300 K, at $I = 35$ mA, showing the droplet creation and annihilation fields.}}
	\label{MR}
\end{figure}

Figure~\ref{MR} shows MR curves at different temperatures at a fix current of 35 mA. As the field is swept up from 0, the droplet nucleates at small field values (it is difficult to characterize because there are also excitations at very small fields related to magnetic domains in the FL); when the field is further increased, the STT effect is no longer able to overcome the dissipation at higher applied field and the droplet annihilates. Sweeping the field back to zero creates the droplet state at lower field values compared with the annihilation of the sweep up, showing hysteresis, and in  continuing to lower fields it annihilates as the STT effect is too small to sustain the droplet state.
For the sake of simplicity, our analysis is only of the annihilation of the droplet when sweeping the field up, and creation when sweeping it down, i.e. in the high-field range, and we will refer to the value of these fields as annihilation and creation fields, respectively. In the following we report only the case of one polarity of the applied field, as the results with the other polarity are analogous.

\begin{figure}
	\includegraphics[width=0.7\columnwidth]{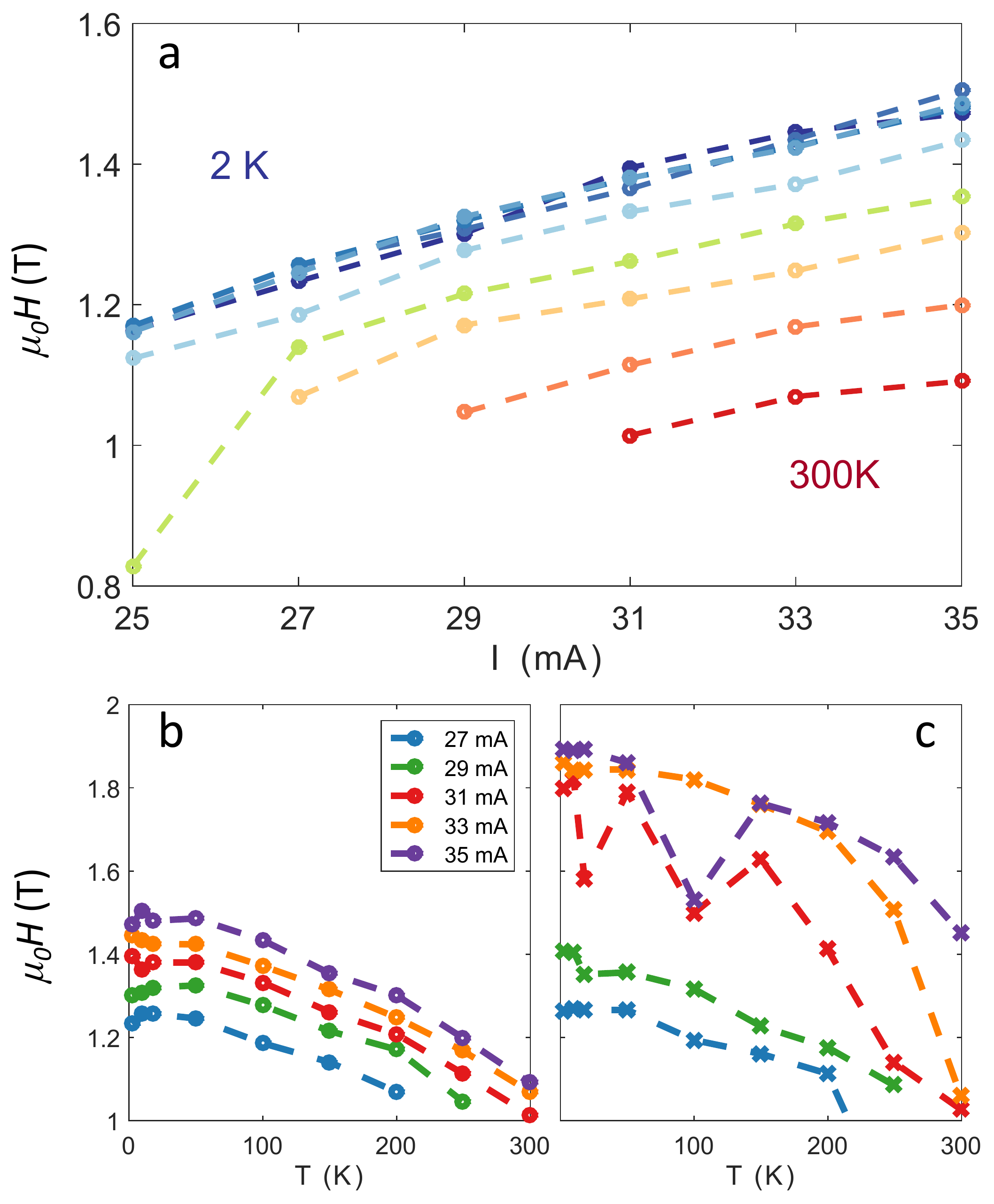}
	\caption{(a) Experimental stability map of soliton states as a function of temperatures (50 K to 300 K). Colors from dark blue, 2 K, to dark red, 300 K indicate the temperatures at which the sample was measured (2, 10, 18, 50, 100, 150, 200, 250, and 300 K). We have only plotted the creation points when sweeping the field down. (b) shows the same data as in (a) but plotted as a function of T for different applied currents and (c) is the same plot for the annihilation fields that are in equal or larger, when there is hysteresis.}
	\label{fig:stab}
\end{figure}
A stability diagram, with measurements in the temperature range from 2 K to 300 K, is shown in Fig.~\ref{fig:stab}a. The curves separate the droplet state (at lower fields and larger currents) from the non-droplet state (at large fields and low currents). Colors from dark blue, corresponding to 2 K, to dark red, corresponding to 300 K indicate the temperature at which the sample was measured. We have only plotted the creation points when sweeping the field from a large value down to zero because the values of the annihilation fields (when sweeping from 0 to a large field value) have a broader distribution and thus the line separating one state from the other is not well defined. We also omit the onset and annihilation at lower fields caused by the canting of the polarizing layer into the film plane and the associated reduction in the perpendicular component of the spin polarization. Figure\ \ref{fig:stab} shows the field values at fixed currents for both the onset (\ref{fig:stab}b) of droplet states (the same used to create the main stability diagram)  and the annihilation (\ref{fig:stab}c). We notice here that Fig.\ \ref{fig:stab}c shows that the values for annihilation have an uncertainty that is not visible in the values corresponding to the onset indicating that the onset and annihilation might have a different origin.

\begin{figure}
	\includegraphics[width=0.9\columnwidth]{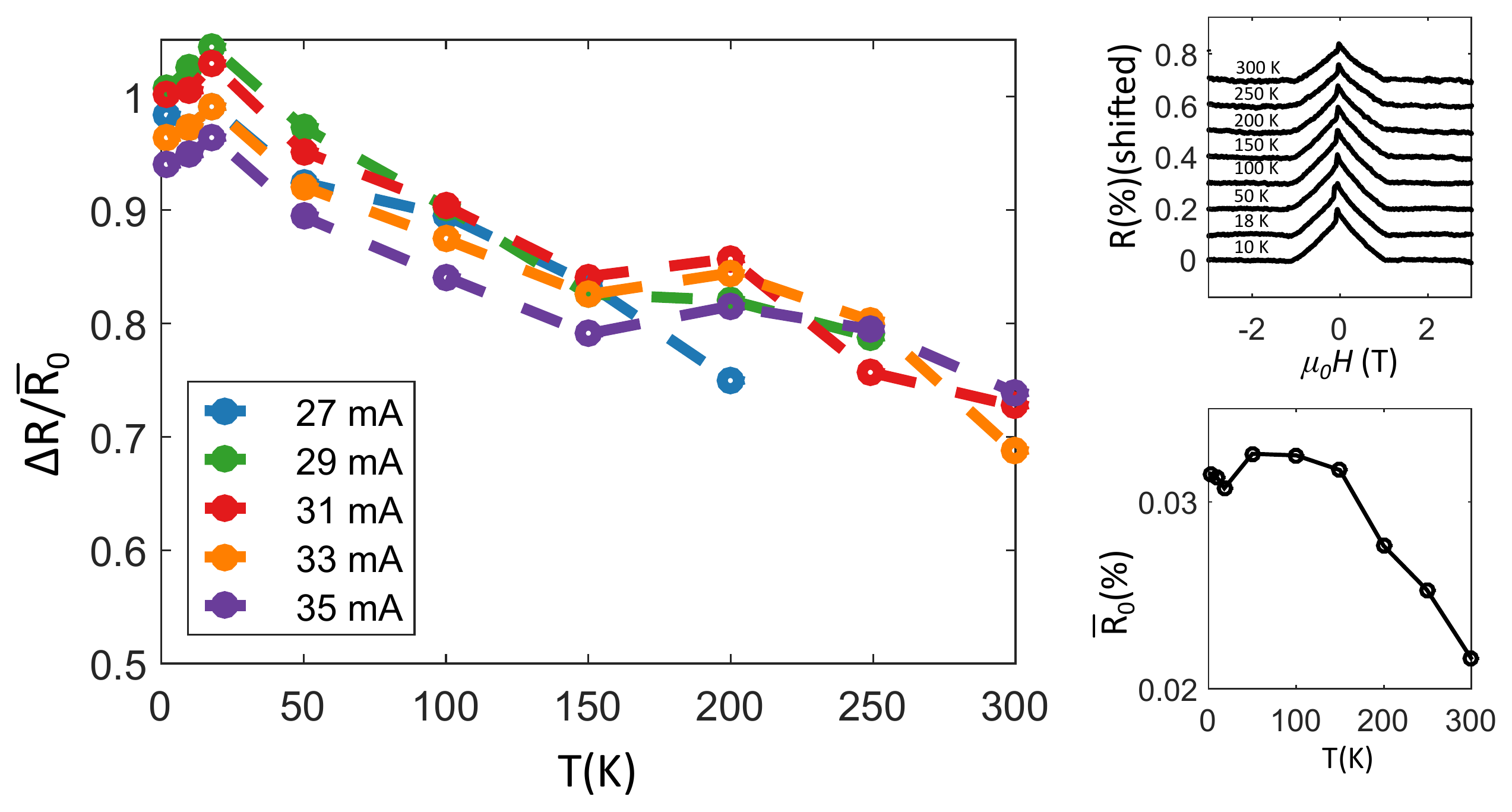}
	\caption{(Left-hand-side panel) Normalized height of the jump $\Delta R/\overline{R}_0$ as a function of temperature for current values of 27 mA (blue), 29 mA (green), 31 mA (red), 33 mA (orange), and 35mA (purple). (Top right-hand-side panel) MR measurements as the magnetic field is swept down from 3T to -3T, for different temperatures in the range (2 - 300) K, using an applied current of $I = -2$ mA. (Bottom right-hand-side panel) Evolution of the overall change in magnetoresistance with temperature $\overline{R}_0$.}
	\label{fig:mrvst}
\end{figure}

In order to evaluate the change in the characteristic resistance associated with the droplet state at different temperatures we must consider the change in the nanocontacts intrinsic MR as a function of temperature. Thus we measured MR curves at a current value of -2 mA. The sign of the current was chosen to avoid excitations in the FL, i.e. it favors damping and stabilizes the FL magnetization. Moreover, a small value of $|2|$ mA minimizes the
STT, yet allows for a measurement of the resistance.  The top right-hand-side panel of Fig.\ \ref{fig:mrvst} shows the measured MR curves at different temperatures. The step at small negative values visible in the MR curves is associated with the reversal of the FL in the direction of the external field, corresponding to the anisotropy field (the curves are measured by decreasing the magnetic field). From the MR curves we calculated the overall change in resistance, $\overline{R}_0$, at each temperature, which are plotted in the lower right-hand-side panel of Fig.\ref{fig:mrvst}. The variations in resistance coming from the change in the droplet state are thus normalized by the intrinsic MR value at that particular field value in order to remove any temperature dependence of the MR. The left-hand-side panel of Fig.\ \ref{fig:mrvst} shows the temperature dependence of the normalized variation in resistance. A value of 1 would mean that the FL and PL magnetizations are antiparallel. As temperature increases, the reversal decreases from almost full reversal at 10-50 K to approximately 70 $\%$ at room temperature.

\newpage
\section*{Curve fitting}

\begin{figure}[htp]
	\includegraphics[width=0.9\columnwidth]{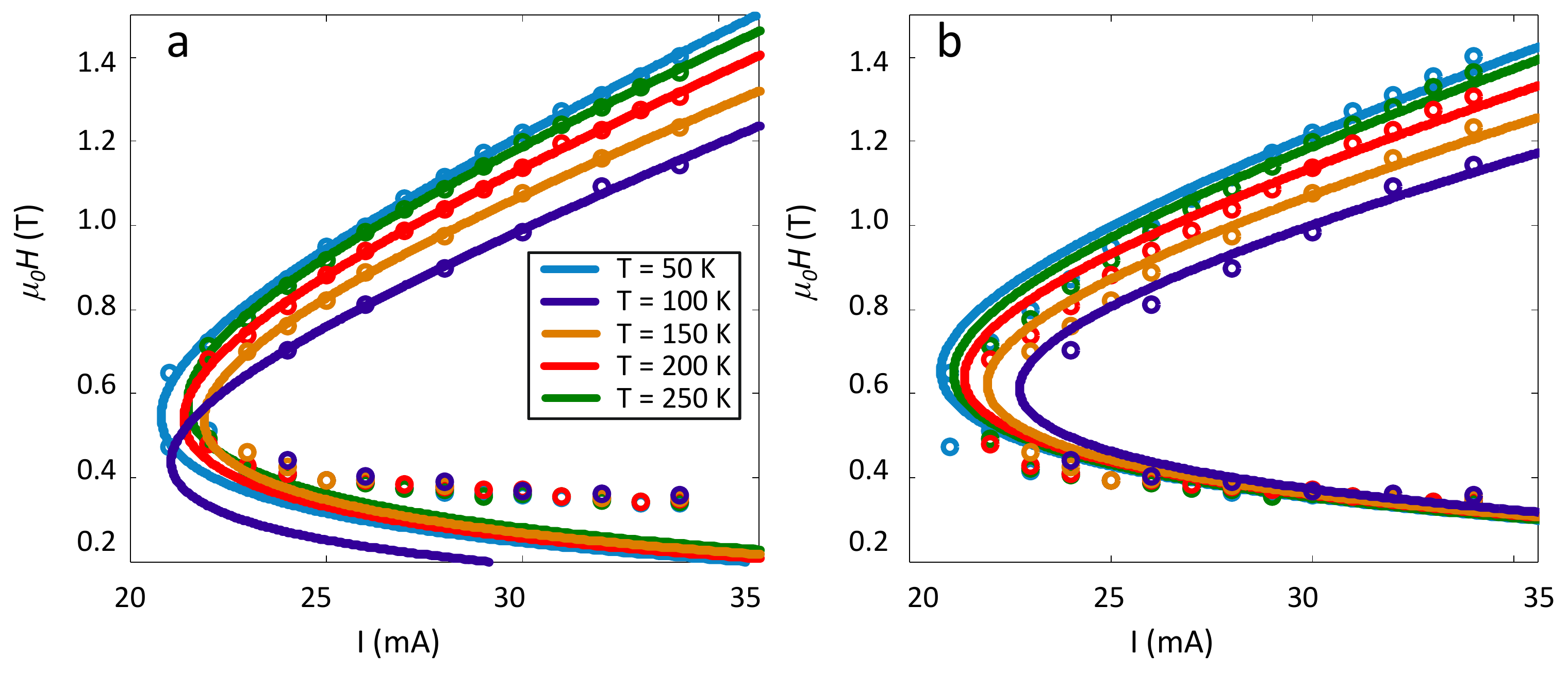}
	\caption{Fits to the stability maps of soliton states as a function of temperatures (50 K to 250 K). The points show measured data. The curves are fits to the data. The fit in (a) uses all the data points whereas the fit in (b) uses only the points that represent the onset of a droplet soliton when sweeping the field from high to low values ($\mu_0H>0.6$ T)}
	\label{fignofit}
\end{figure}

The expression suggested by Chung \emph{et al.} \cite{Akerman2016NatComm} for the the onset current of the droplet soliton that accounts for the effect of a PL not aligned with the FL does not fit our data.
Figure~ \ref{fignofit} shows the fitting of the data presented in the main manuscript using the expression
\begin{equation}
I=a(\mu_0H)+\frac{b}{\mu_0H}+c,
\label{nofits}
\end{equation}
where $a$, $b$, and $c$ are the fitting parameters.
We present two curve fits considering either all the points (in Fig. \ref{fignofit}a) or only the points that represent the onset of a droplet soliton when sweeping the field from high to low values ($\mu_0H>0.6$ T). The fits are poor and cannot capture the behavior at both large and small fields.
\\

We can fit the data by considering a shift in the applied field, $H-H_0$ instead of $H$, of approximately 0.27 T.

Next we analyze the evolution of the fitting parameters in the equation
\begin{equation}
I=a\mu_0(H-H_0)+\frac{b}{\mu_0(H-H_0)}+c,
\label{fits}
\end{equation}
as a function of temperature.

\begin{figure}[htp]
	\includegraphics[width=0.8\columnwidth]{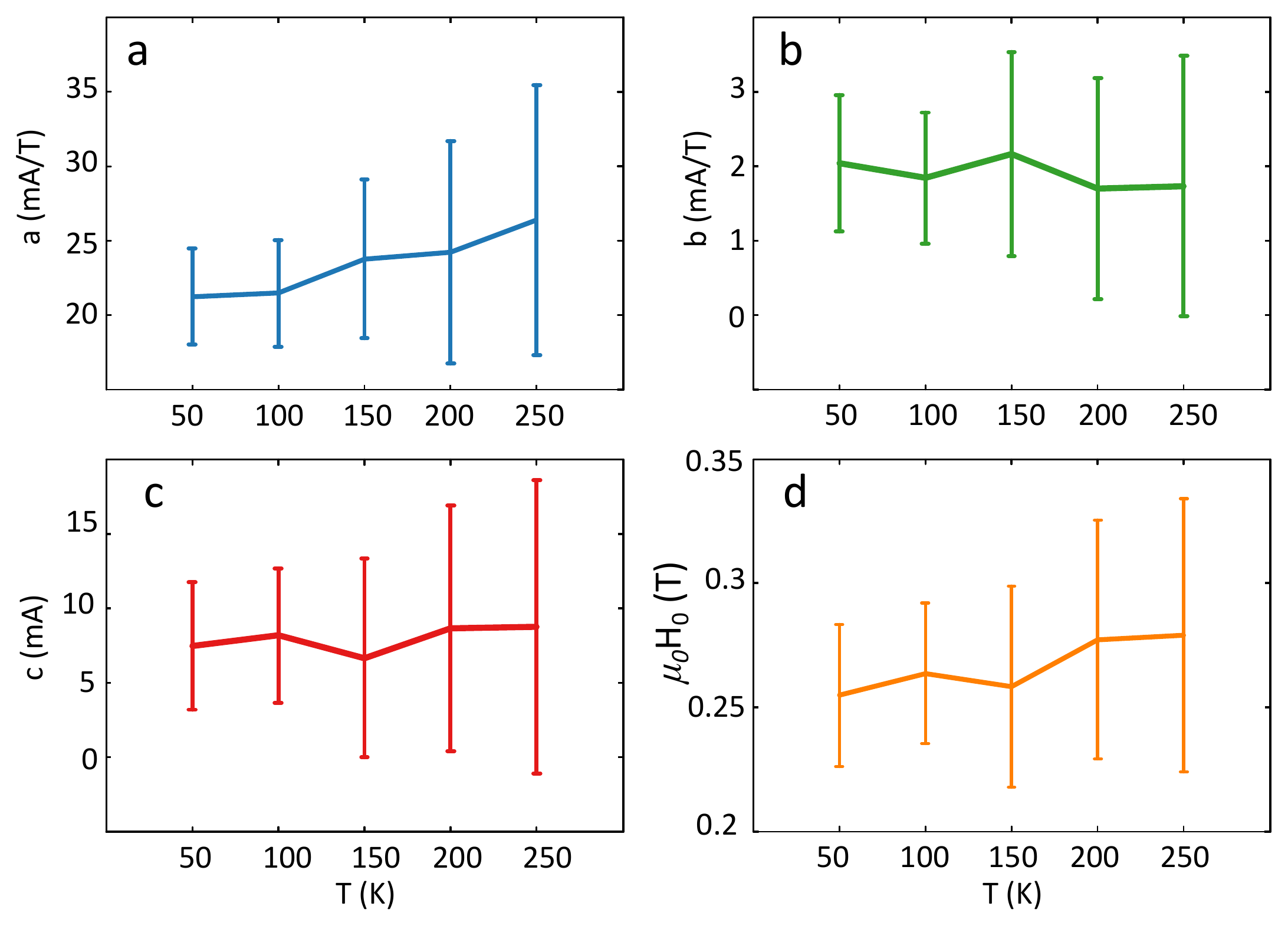}
	\caption{Temperature evolution of free parameters used in Eq.\ \ref{fits}}
	\label{fig:freepars}
\end{figure}

We first considered all four parameters as free parameters. The values are summarized in Table \ref{tab:freepars}, and the same data is plotted in Fig.~\ref{fig:freepars}.

\begin{table}[htp]
	\caption{Evolution of four free parameters}
	\begin{center}
		\begin{tabular}{c|c|c|c|c}
			$T$ (K) & $a$ (mA/T) & $b$ (mA$\cdot$T) & $c$ (mA) & $\mu_0H_0$ (T) \\ \hline
			50 & ~$21 \pm 3$~ & ~$2.0 \pm 0.9$~ & ~$7 \pm 5$~ & ~$0.25 \pm 0.03$ \\
			100 & ~$21 \pm 4$~ & ~$1.8 \pm 0.8$~ & ~$8 \pm 5$~ & ~$0.26 \pm 0.03$ \\
			150 & ~$24 \pm 5$~ & ~$2.2 \pm 1.3$~ & ~$7 \pm 7$~ & ~$0.26 \pm 0.04$ \\
			200 & ~$24 \pm 7$~ & ~$1.7 \pm 1.5$~ & ~$9 \pm 9$~ & ~$0.28 \pm 0.05$ \\
			250 & ~$26 \pm 9$~ & ~$1.7 \pm 1.8$~ & ~$10 \pm 11$~ & ~$0.28 \pm 0.06$
		\end{tabular}
	\end{center}
	\label{tab:freepars}
\end{table}%

\begin{figure}[htp]
	\includegraphics[width=0.8\columnwidth]{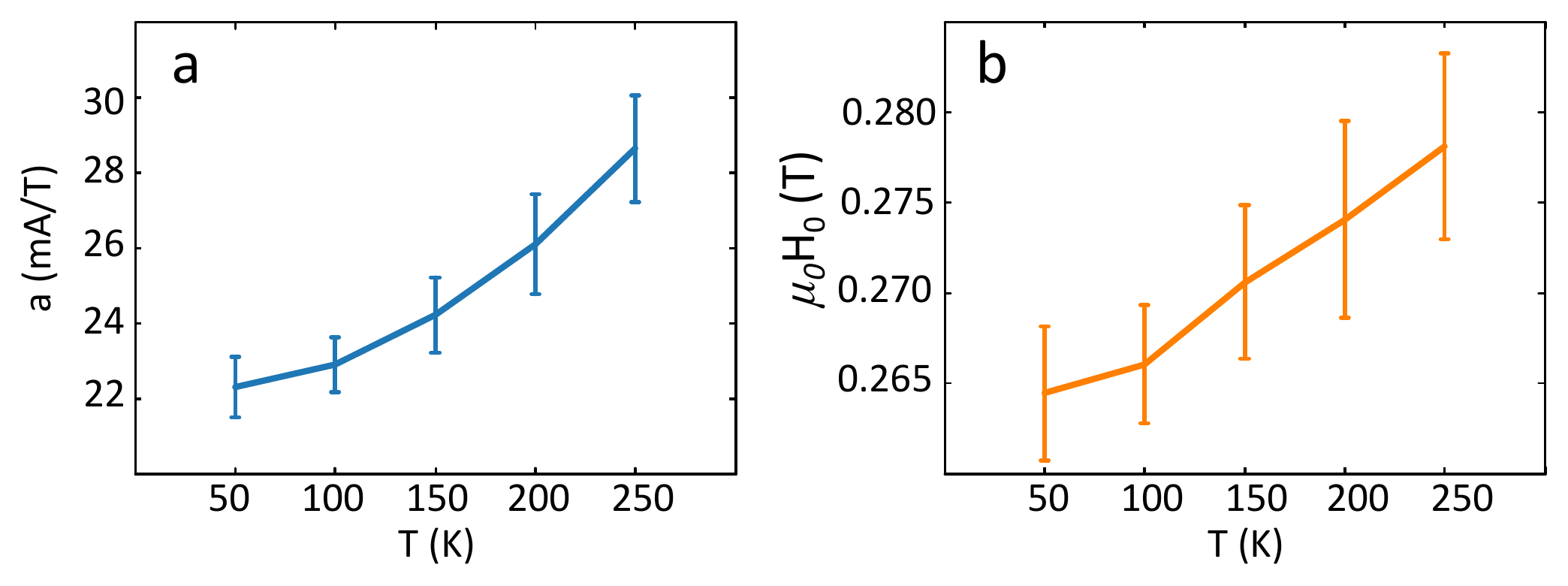}
	\caption{Evolution of parameters, fixing $b = 1.9$~mA$\cdot$T and $c = 7$~mA}
	\label{fig:fixed_bc}
\end{figure}

We observe that the parameter $b$ is almost constant (this parameter only matters at low fields) and the parameter $c$ has also a small variation in temperature. Thus, we fixed $b = 1.9$~mA$\cdot$T and $c = 1.7$~mA. The parameters are again summarized in Table \ref{tab:fixed_bc}, and the same data are plotted in Fig.~\ref{fig:fixed_bc}.

\begin{table}[htp]
	\caption{Evolution of parameters, fixing $b = 1.9$~mA$\cdot$T and $c = 7$~mA}
	\begin{center}
		\begin{tabular}{c|c|c}
			$T$ (K) & $a$ (mA/T) & $\mu_0H_0$ (T) \\ \hline
			50 & ~$22 \pm 0.8$~ & ~$0.264 \pm 0.004$ \\
			100 & ~$23 \pm 0.7$~ & ~$0.266 \pm 0.003$ \\
			150 & ~$24 \pm 0.9$~ & ~ $0.271 \pm 0.004$ \\
			200 & ~$26 \pm 1.3$~ & ~ $0.274 \pm 0.005$ \\
			250 & ~$29 \pm 1.4$~ & ~ $0.278 \pm 0.005$
		\end{tabular}
	\end{center}
	\label{tab:fixed_bc}
\end{table}%

\newpage

We observe that the parameter $a$, which indicates the slope of the line of the onset current (should be proportional to dissipation), increases considerably from 22 mA/T to almost 30 mA/T. Additionally the parameter $h_0$ that matters mostly at low fields also increases about a 10 \% with temperature.

\end{document}